\documentclass[11pt,a4paper]{article}
\usepackage{amsmath}
\usepackage{amssymb}
\usepackage{cite}
\usepackage{epsfig}
\usepackage{feynmp}
\usepackage{xspace}
\usepackage{flafter}
\usepackage{graphicx}

\newcommand{\Ca}{\ensuremath{C_{\!A}}}
\newcommand{\Cf}{\ensuremath{C_{\!F}}}
\newcommand{\Nc}{\ensuremath{N_{\!C}}}

\newcommand{\as}{\ensuremath{\alpha_s}\xspace}
\newcommand{\gs}{\ensuremath{g_s}}

\title{\vspace*{-3.cm}
\begin{normalsize}
\begin{flushright}
CERN-PH-TH/2010-003\\
KA-TP-05-2010
\end{flushright}
\end{normalsize}
\vspace*{1cm}
Azimuthal Angle Correlations for Higgs Boson plus Multi-Jet Events
\author{Jeppe R.~Andersen$^{a}$,
  Ken Arnold$^{b}$, Dieter Zeppenfeld$^{b}$\\\mbox{}\\
  $^a$Theory Division, Physics
  Department, CERN, CH-1211 Geneva 23, Switzerland\\
  $^b$Institute for Theoretical Physics, Karlsruhe Institute of Technology, 
  D-76128 Karlsruhe}
}
\begin{document}
\maketitle

\begin{abstract}
  At lowest order in perturbation theory, the scattering matrix element for
  Higgs boson production in association with dijets displays a strong
  correlation in the azimuthal angle between the dijets, induced by the
  $CP$-properties of the Higgs Boson coupling. However, the phase space cuts
  necessary for a clean extraction of the $CP$-properties simultaneously
  induce large corrections from emissions of hard radiation and thus
  formation of extra jets. The current study concerns the generalization of
  $CP$-studies using the azimuthal angle between dijets beyond tree-level and
  to events with more than just two jets. By analyzing the High Energy Limit of
  hard scattering matrix elements we arrive at a set of cuts optimized to
  enhance the correlation, while maintaining a large cross section, and an
  observable, which is very stable against higher order corrections. We
  contrast the description of Higgs boson production in association with jets
  at different levels: for tree-level $hjj$ and $hjjj$ matrix elements, 
  for $hjj$ matrix elements plus parton shower, and in a
  recent all-order framework, which converges to the full, all-order
  perturbative result in the limit of large invariant mass between all
  produced particles.
\end{abstract}

\section{Introduction}
\label{sec:introduction}
One of the prime goals of experiments at the CERN 
Large Hadron Collider (LHC) is the search for the Higgs boson(s) 
which, within the Standard Model (SM) and many of its extensions,
provide direct access to the dynamics of electroweak symmetry breaking.
Once discovered, the focus of Higgs physics will turn to the study of 
Higgs boson properties, like its mass, spin, \ensuremath{CP} parity and the strength and
structure of Higgs boson couplings to heavy fermions and gauge bosons.

Among the various Higgs channels at the LHC, the production of a Higgs boson 
in association with two energetic jets has emerged as particularly promising
in providing information on the dynamics of the Higgs sector. For a SM-like
Higgs boson, the vector boson fusion process, $qq\to qqh$, is expected
to provide Higgs signals in the $h\to W^+W^-$~\cite{Rainwater:1999sd}, 
$h\to \tau^+\tau^-$~\cite{Rainwater:1998kj,Plehn:1999xi}, 
and/or $h\to \gamma\gamma$~\cite{Rainwater:1997dg} 
decay channels~\cite{Asai:2004ws,Ball:2007zza}, 
depending on the Higgs boson mass, 
and produce crucial information for extracting the size of Higgs boson 
couplings~\cite{Zeppenfeld:2000td,Duhrssen:2004cv,Lafaye:2009vr}.
A second important source of Higgs plus 2-jet events are gluon fusion
processes, such as $qq\to qqh$, $gq\to gqh$ or 
$gg\to ggh$~\cite{Kauffman:1996ix}.  
For favorable values of the Higgs boson mass, gluon fusion induced 
$hjj$ events should be visible at the LHC
in the $h\to W^+W^-$~\cite{Klamke:2007cu} and 
$h\to \tau^+\tau^-$~\cite{klamke08,klamke09} channels. 

Even in the presence of substantial SM backgrounds, the azimuthal angle 
correlations of gluon fusion induced $hjj$ events can be used for 
establishing \ensuremath{CP} properties of the interactions of the Higgs 
boson~\cite{Klamke:2007cu,klamke08,klamke09}. In a SM-like situation, 
the Higgs interaction with gluons is mostly mediated by a top-quark 
loop. For a Higgs boson 
which is lighter than the top quark, the resulting $hjj$ cross section can
be determined to good approximation by an effective Lagrangian of energy 
dimension five, which is given by~\cite{Kauffman:1996ix,Kauffman:1998yg}
\begin{equation}
{\cal L}_{\rm eff} =
\frac{y_t}{y_t^{SM}}\cdot\frac{\alpha_s}{12\pi v} \cdot h \,
G_{\mu\nu}^a\,G^{a\,\mu\nu} +
\frac{\tilde y_t}{y_t^{SM}}\cdot\frac{\alpha_s}{8\pi v} \cdot A \,
G^{a}_{\mu\nu}\,\tilde{G}^{a\, \mu\nu}\;,
\label{eq:ggS}
\end{equation}
where $G^{a}_{\mu\nu}$ denotes the gluon field strength and
$\tilde{G}^{a\, \mu\nu} = 1/2\, 
G^{a}_{\rho\sigma}\,\varepsilon^{\mu\nu\rho\sigma}$ its dual. 
$y_t^{SM}=m_t/v$ is the SM Yukawa coupling of the Higgs boson to 
top quarks. The two terms result from a $\overline{t} th$ and a 
$\overline{t}i\gamma_5 tA$
coupling of the (pseudo)scalar Higgs, respectively, and they lead to 
distinctively different distributions of the azimuthal angle between the 
two jets: the \ensuremath{CP}-even $hgg$ coupling produces a minimum for 
$\phi_{jj}=\pm\pi/2$ while a $CP$-odd $Agg$ coupling leads to
minima at $\phi_{jj}=0$ and $\pm\pi$. The azimuthal angle modulations 
get particularly pronounced when the two jets are widely separated in
rapidity. Equivalent effects are expected in 
vector boson fusion and have been discussed 
in~\cite{Plehn:2001nj,Hankele:2006ma} for the idealized situation of 
parton level events at leading order (LO).

For a more realistic simulation of $hjj$ events, effects from multiple 
parton emission must be included, i.e. a full parton shower analysis with
subsequent hadronization and jet reconstruction should be performed. There 
are two effects which may substantially reduce the azimuthal angle 
correlations predicted with LO matrix elements: (i) the emission of 
additional gluons between the widely separated jets may lead to 
a decorrelation of azimuthal angle distributions and 
(ii) even the very definition of the two tagging jets becomes ambiguous 
in the multi-jet environment which is expected after parton shower.

A first analysis of $hjj$ events with full parton shower simulation was 
performed by Odagiri~\cite{Odagiri:2002nd}, 
which, however, was built on LO $gg\to h$ 
hard matrix elements which do not provide the full dynamics of non-trivial
azimuthal angle correlations of the produced partons. A subsequent {\sc Herwig} 
analysis with parton shower corrections to hard matrix elements 
for LO Higgs + 2 and Higgs + 3 parton production~\cite{DelDuca:2006hk}
indicated that parton
shower effects do lead to some deterioration of the azimuthal angle 
correlations of the two leading jets (defined as the highest transverse 
momentum jets). The observed deterioration was larger, however, than 
what was observed in a full NLO calculation of $hjj$ production in 
the effective Lagrangian approximation~\cite{Campbell:2006xx}. 
This discrepancy suggests 
that a parton shower approach generically leads to an underestimate of 
jet azimuthal angle correlations. The most likely cause is that the 
parton shower generates radiation which is basically flat in 
azimuthal angle and, hence, all events, where one of the tagging jets arises
purely from the parton shower, have lost the azimuthal correlation. In addition,
hard radiation can lead to substantial changes in the angle of 
the selected tagging jet which results in further decorrelation.

The full complexity of the jets in terms of the particles arising from the
shower initiated by a few hard partons can only be described 
within a general purpose
Monte Carlo\cite{Bahr:2008pv,Sjostrand:2007gs,Gleisberg:2008ta} with a
hadronization model. However, the description reached in such models by
starting from soft- and collinear approximations may not be satisfactory for
the \emph{hard} (in transverse momentum), \emph{perturbative} corrections;
the description of the hard radiation can be repaired by
matching\cite{Lonnblad:1992tz,Catani:2001cc,Mangano:2006rw} to fixed order
calculations; but this means that the effects of such hard emissions are not
resummed, and furthermore, virtual corrections (and the resulting weighting
of samples with varying jet multiplicity) are estimated using only the
Sudakov factor from the shower. When the interest is in the number of jets
and their topology, rather than the description of the constituents of each
jet, then the hard, multi-parton matrix elements can be approximated to all
orders, based on the factorization properties of the scattering matrix
element in the limit of large invariant mass between each hard particle,
known as the \emph{High Energy Limit} or
\emph{Multi-Regge-Kinematics}\cite{Fadin:1975cb,Kuraev:1976ge,Kuraev:1977fs}. These
ideas were developed further in
Ref.~\cite{Andersen:2008ue,Andersen:2008gc,Andersen:2009nu,Andersen:2009he},
where a partonic Monte Carlo program was constructed, which captures to all
orders the leading logarithmic behavior for emissions under large invariant
mass.

The goal of the present paper is two-fold. First, we present an improved 
observable, replacing the azimuthal angle between the leading jets as a
probe for the \ensuremath{CP}-properties of the Higgs coupling to heavy quarks.
The improvements are then assessed by simulating Higgs events in 
association with two or more jets at different levels of complexity.
In Section~\ref{sec:lessons-from-he}
we first discuss how insight into the structure of the
perturbative corrections in the High Energy or \emph{Multi-Regge Kinematic}
(MRK) limit can assist in designing observables, which extract the relevant
kinematic information of the Higgs Boson vertex and are stable against higher
order corrections. The resulting redefinition of the azimuthal angle between
the two final state partons, in terms of the angle between jet clusters,
is then first probed in Section~\ref{sec:fixed-order-studies}, where we 
compare results for full fixed order matrix elements: the tree-level 
calculations for $hjj$ and $hjjj$ production are juxtaposed and the idealized
final states in the two- and three-parton configurations are used for deriving
upper bounds on the azimuthal angle correlations which can be expected in a 
full simulation. For this full simulation we compare two approaches in 
Section~\ref{sec:corr-mutli-jet}: the conventional parton shower  
is generated with {\sc Herwig++}~\cite{Bahr:2008pv} while the 
{\it Multi-Regge Kinematics} is simulated with the programs developed in 
Refs.~\cite{Andersen:2008ue,Andersen:2008gc,Andersen:2009nu,Andersen:2009he}. 
A summary and final conclusions are drawn in Section~\ref{sec:conclusions}.

\section{Lessons From the High Energy Limit}
\label{sec:lessons-from-he}
In this section we will discuss how the insight into the structure of the
perturbative corrections in the 
MRK limit can assist in designing observables, which extract the relevant
kinematic information of the Higgs Boson vertex and are stable against higher
order corrections. 

The MRK limit of the $2\to n$ scattering process is characterized by a large
invariant mass between each of the produced particles, each of a fixed,
perturbative transverse momentum. Specifically, for the scattering resulting
in jets with momenta $p_1,\cdots, p_n$, the MRK limit is given by
\begin{align}
  \label{eq:MRKlimit}
  \begin{split}
    \forall i\in \{2,\ldots, n-1\}: y_{i-1}\gg y_i \gg y_{i+1}\\
    \forall i,j: |\mathbf{p}_{i\perp}|\approx |\mathbf{p}_{j\perp}|,\quad 
    |\mathbf{q}_{i\perp}|\approx |\mathbf{q}_{j\perp}|.
  \end{split}
\end{align}
Here $\mathbf{q}_{i\perp}$ is the transverse part of the 
momentum of the $i^\mathrm{th}$ $t$-channel
propagator, defined as $q_i=p_a-\sum_{j=1}^i p_j$.

In this extreme kinematic limit, the hard scattering matrix element
simplifies for two reasons:
\begin{enumerate}
\item The contribution to jet production is dominated by the partonic
  channels which allow a color octet exchange between all pairs of
  neighboring particles in rapidity. Furthermore, in the MRK limit all such
  partonic channels have a universal behavior, dictated by the residues in
  $t$-channel momenta (see Ref.\cite{Andersen:2009nu,Andersen:2009he} for
  more details).
\item The kinematic invariants relevant for the description of the dominant
  part of the scattering amplitude simplify in the MRK limit, resulting in a
  dependence on transverse momenta only.
\end{enumerate}
Recent
efforts\cite{Andersen:2008ue,Andersen:2008gc,Andersen:2009nu,Andersen:2009he}
have concentrated on developing a formalism exploiting the universality of
the dominance of $t$-channel poles (point 1), without introducing unnecessary
kinematic approximations, thus obtaining a formalism which can be applied
with good results away from the strict MRK limit. However, the discussion in
the remainder of this section will be based on applying the full kinematic
approximations of this limit.

In the MRK limit, the square of the tree-level hard scattering matrix element
takes the following form for Higgs boson production in association with $n$
gluon jets, when the rapidity of the Higgs boson is in-between jets $j$ and
$j+1$:-
\begin{align}
  \label{eq:ngluonplush}
  \left| \mathcal{M}_{gg\to g\cdots ghg\cdots g}\right |^2 \rightarrow \frac {4 {\hat s} ^2}
    {\Nc^2-1}\ 
   \left( \prod_{i=1}^j\frac{\Ca\ \gs^2}{\mathbf{p}_{i\perp}^2} \right)
   \frac{|C^H(\mathbf{q}_{a\perp}, \mathbf{q}_{b\perp})|^2}{\mathbf{q}_{a\perp}^2\  \mathbf{q}_{b\perp}^2}\
   \left(\prod_{i=j+1}^n\frac{\Ca\ \gs^2}{\mathbf{p}_{i\perp}^2}\right),
\end{align}
where $\mathbf{q}_{a\perp}=-\sum_{i=1}^j\ \mathbf{p}_{i\perp}$, where $j$ is the number of gluons
with rapidity smaller than that of the Higgs boson, and
$\mathbf{q}_{b\perp}=\mathbf{q}_{a\perp}-\mathbf{p}_{h\perp}$. In this limit, the contribution from
quark-initiated processes is found by just a change of one color factor
$\Ca\to\Cf$ for each incoming gluon replaced by a quark. The effective
vertex for the coupling of the Higgs boson to two off-shell gluons through a
top loop is in the combined large-$m_t$ and MRK limit\cite{DelDuca:2003ba}
\begin{align}
  \begin{split}
    \label{eq:Ch}
    C^H(\mathbf{q}_{a\perp},\mathbf{q}_{b\perp})\ &=\ i\ \frac A 2\ \left(|\mathbf{p}_{h\perp}|^2 -
      |\mathbf{q}_{a\perp}|^2 - |\mathbf{q}_{b\perp}|^2 \right),\\
    A\ &=\frac{y_t}{y_t^{SM}}\ \frac {\as} {3\pi v},\quad v=246\ \mbox{GeV}.
  \end{split}
\end{align}
Since $p_h=q_a-q_b$, $C^H$ is given by
\begin{align}
  \label{eq:Chcosthetaab}
  C^H(\mathbf{q}_{a\perp},\mathbf{q}_{b\perp})\ &=\   
  -i\ A\ |\mathbf{q}_{a\perp}|\ |\mathbf{q}_{b\perp}|\ 
  \cos (\phi_{\mathbf{q}_{a\perp} \mathbf{q}_{b\perp}}) = 
  -i\ A\ \mathbf{q}_{a\perp}\cdot \mathbf{q}_{b\perp}
\end{align}
In the case of a $CP$-odd coupling of a pseudo-scalar to the fermions in the
loop, the high-energy factorization properties and the formula in
Eq.~\eqref{eq:ngluonplush} still hold, with $C^H$ replaced by
\begin{align}
  \label{eq:CAcosthetaab}
  C^A(\mathbf{q}_{a\perp},\mathbf{q}_{b\perp})\ &=\ 
  i\ B\ |\mathbf{q}_{a\perp}|\ |\mathbf{q}_{b\perp}|\ 
  \sin (\phi_{\mathbf{q}_{a\perp} \mathbf{q}_{b\perp}}) = 
  i\ B\ (\mathbf{q}_{a\perp}\times\mathbf{q}_{b\perp})\cdot \hat{z},
\end{align}
where $\hat{z}$ is a unit vector along the beam axis whose sign is given by 
the $z$ component of $q_a-q_b$. In the notation of Ref.~\cite{Klamke:2007cu}, 
\begin{align}
  B=\frac{\tilde y_t}{y_t^{SM}}\ \cdot\ \frac{\alpha_s}{2\pi v}.
\end{align}
The $CP$-properties of the coupling of the (pseudo-) scalar are reflected in
the dependence on the azimuthal angle between $q_a$ and $q_b$ of the
effective vertex for the coupling to gluons via a fermion loop (e.g.~$C^H$ or
$C^A$).  It is this dependence that the observables and cuts should
emphasize. In the case of Higgs boson production in association with just two
jets, $\mathbf{q}_{a\perp},\mathbf{q}_{b\perp}$ are (up to a sign) given by
the transverse momentum of the jets (at lowest order in perturbation theory),
and thus the $CP$-properties of the Higgs boson coupling can be extracted by
studying the azimuthal angle between the two jets\cite{Plehn:2001nj}.

Eq.~\eqref{eq:ngluonplush} is valid for the rapidities of the Higgs boson
within the rapidity interval spanned by the partons. In fact, the 
factorization of the amplitude into a part which depends on the momenta 
of the partons with
rapidities much smaller than that of the Higgs boson, a Higgs boson vertex,
and a function depending on the momenta of partons with rapidities much
larger than that of the Higgs boson is valid also when no requirement is
placed on the invariant mass between the partons on either side of the Higgs
boson\cite{DelDuca:1999ha}. Therefore, we will divide the observable jets
into two groups and require that all jets are well separated from the
Higgs boson direction,
\begin{align}
  \begin{split}
  \label{eq:cuts}
    &\exists j_a : y_{j_a}< y_{h},\quad \exists j_b : y_{j_b}> y_{h},\\
    &\forall j\in\{ \mbox{jets} \}\ :\ |y_j-y_h|>y_{\mbox{sep}}
  \end{split}
\end{align}
In the following we always use the $k_t$ jet algorithm as implemented in
Ref.\cite{Cacciari:2005hq} with $R=0.6$ and investigate jets with a
transverse momentum greater than 40~GeV. 

While it is clear that in LO $hjj$ production the \ensuremath{CP}-structure 
of the Higgs boson vertex can be revealed by studying the azimuthal angle 
between the two jets, the question arises which angle to study in events 
with more than two jets. Traditionally, the observable applied to extract  
the \ensuremath{CP} structure has been the azimuthal angle between the 
hardest jets of the
event~\cite{Hankele:2006ma,Klamke:2007cu,DelDuca:2006hk}, within a set of cuts
based on the rapidity separation between the two hardest jets.
However, Eqs.~\eqref{eq:ngluonplush}-\eqref{eq:Chcosthetaab} suggest
that the \ensuremath{CP} structure is more clearly revealed if instead 
the azimuthal angle between $q_a$ and $q_b$ is studied, with
\begin{align}
  \begin{split}
    \label{eq:qaqb}
    q_a\ &=\ \sum_{j\in \{\mbox{jets} : y_j<y_h\}} p_j,\qquad \qquad
    q_b\ =\ \sum_{j\in \{\mbox{jets}  : y_j>y_h\}} p_j,\\
    \phi_2\ &=\ \angle(\mathbf{q}_{a\perp},\mathbf{q}_{b\perp})
  \end{split}
\end{align}
In the MRK limit, $(1/\sigma\ d\sigma/d\phi_2)$ is stable against higher
order corrections, in so far as soft radiation, which is not picked up as jets,
does not impact the angular distribution.

\boldmath
\section{Fixed Order Studies}
\label{sec:fixed-order-studies}
\subsection{Tree level studies for $hjj$ production}
\unboldmath
We start by studying the tree-level predictions for $hjj$ production
for a 14~TeV
proton-proton collider and a Higgs boson mass of 120~GeV. We use the
central pdf set from MSTW2008\cite{Martin:2009iq}, and initially we use a
minimum set of cuts:
\begin{align}
  \label{eq:minimumcuts}
  p_{j\perp}>40\ \mbox{GeV},\quad y_{ja}<y_h<y_{jb},\quad |y_{j_a,j_b}|<4.5.
\end{align}
The main conclusions of this study are insensitive to the choice of
renormalization and factorization scale. In the $hjj$-analysis we 
set $\as^4\to \as(p_{ja\perp}) \as(p_{jb\perp}) \as^2(m_H)$, with
factorization scales for the pdfs of
$\mu_{f,a}=p_{ja\perp},\mu_{f,b}=p_{jb\perp}$. The matrix elements are
extracted from MadGraph\cite{Alwall:2007st}, where the
$m_t\rightarrow\infty$ limit is used.

The distribution of the cross section wrt.~azimuthal angle between the two
jets is often distilled into a single number $A_\phi$:
\begin{align}
  \label{eq:Aphi}
  A_\phi=\frac{\sigma(|\phi_{j_aj_b}|<\pi/4)-\sigma(\pi/4<|\phi_{j_aj_b}|<3\pi/4)+\sigma(|\phi_{j_aj_b}|>3\pi/4)}{\sigma(|\phi_{j_aj_b}|<\pi/4)+\sigma(\pi/4<|\phi_{j_aj_b}|<3\pi/4)+\sigma(|\phi_{j_aj_b}|>3\pi/4)}.
\end{align}
This obviously gives $-1\leq A_\phi\leq 1$, with $A_\phi=0$ representing no
azimuthal correlation between the jets. If there are no other sources of
angular dependence, then a $C\!P$-even coupling of the Higgs boson to two
off-shell gluons through a top-triangle leads to a positive value for
$A_\phi$, whereas a $C\!P$-odd coupling results in a negative value for
$A_\phi$. 

\begin{figure}[tb]
  \centering
  \epsfig{width=\textwidth,file=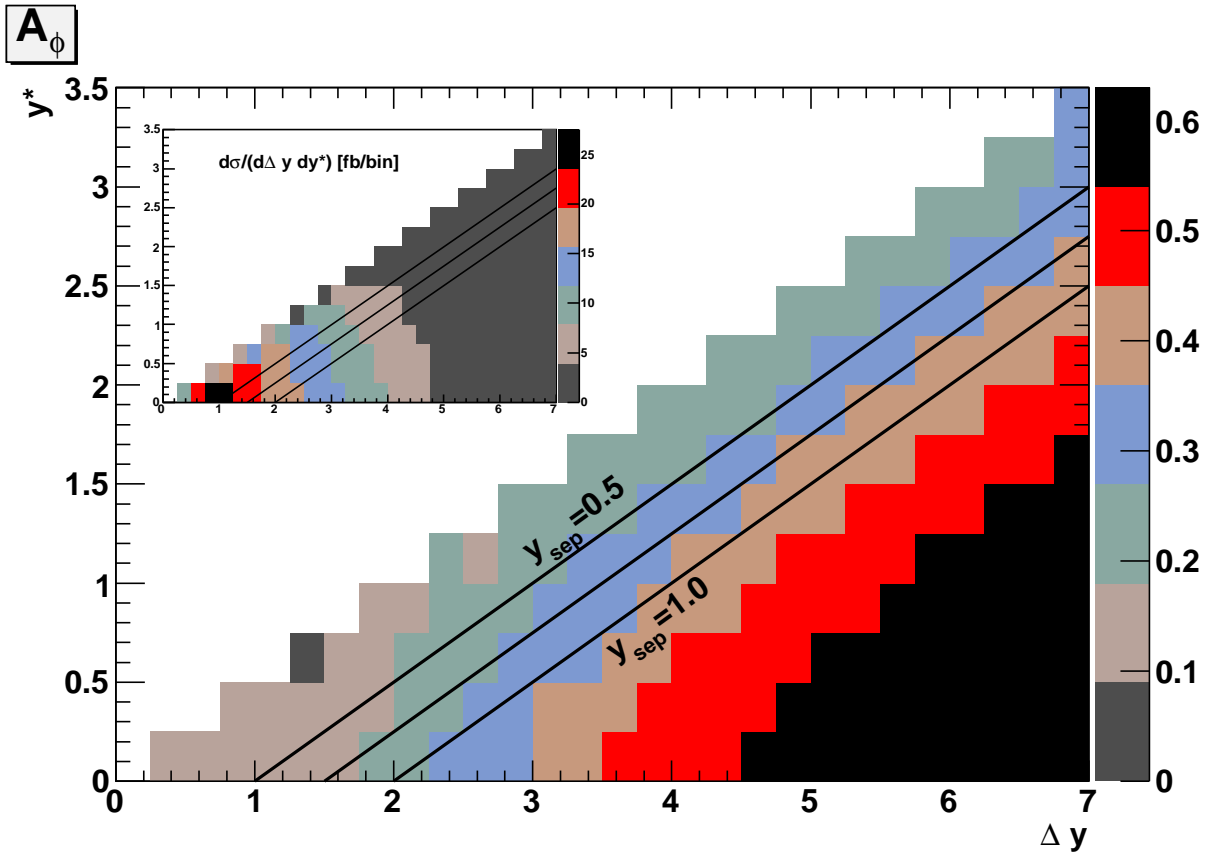}
  \vspace{-6mm}
  \caption{$A_\phi$ vs.~$\Delta y$ and $y^*$ with a bin size of $.25\times
    .25$ units of rapidity, for a SM Higgs boson of mass $m_h=120$~GeV. 
    Insert: $d\sigma/(dy^*d\Delta y)$ in fb/bin.}
  \label{fig:AphivsDyys}
\end{figure}
In Fig.~\ref{fig:AphivsDyys} we plot the value for $A_\phi$ as a function of
$\Delta y$ and $y^*$ given by
\begin{align}
  \label{eq:Deltayys}
  \Delta y= |y_{j_a}-y_{j_b}|,\quad y^*=y_h-\frac{y_{j_a}+y_{j_b}}2.
\end{align}
The bin-size is $.25\times.25$ units of rapidity. The insert shows the cross
section per bin. Because of the requirement of the Higgs boson to be
in-between the jets in rapidity, we have $0<y^*<\Delta y/2$. 
We see that $A_\phi$ is
largest when the jets are separated by a large rapidity interval ($\Delta y$
large), and the Higgs boson is produced near the rapidity-center of the two
jets ($y^*$ small). This is caused by two effects: 
\begin{enumerate}
\item This kinematic setup approaches the MRK limit, where the matrix element
  for the gluon channels have the same kinematic dependence as the
  quark-initiated ones. This dependence is induced by the singularity from a
  gluon exchanged in the $t$-channel, coupling to the Higgs boson field
  through a top triangle. The effects of other gluon-Higgs couplings through
  top loops are suppressed, and only the one giving rise to the strong
  azimuthal correlation survives (see
  Ref.\cite{Andersen:2009nu,Andersen:2009he,Andersen:2008ue,Andersen:2008gc}
  for further details).
\item The azimuthal angular dependence of the $t$-channel propagators is
  suppressed, and the only azimuthal dependence left is induced by the tensor
  structure of $g^* g^* h$-coupling which leads to Eq.~\eqref{eq:Chcosthetaab}.
\end{enumerate}
The insert in Fig.~\ref{fig:AphivsDyys} shows that the cross section is
peaked at $\Delta y\sim 1$ and decreases for increasing $\Delta y$ and
$y^*$. On Fig.~\ref{fig:AphivsDyys} we have also indicated the limits 
in the $(\Delta y,y^*)$-plane corresponding to the three values 
$y_\mathrm{sep}=0.5,0.75,1$, i.e. the lines correspond to
\begin{align}
  \label{eq:ysep}
  \min\left(|y_h-y_{j_a}|,|y_h-y_{j_b}|\right) = y_\mathrm{sep} \, .
\end{align}
Good analyzing power (large $A_\phi$) and a large rate are 
simultaneously retained by imposing a $y_\mathrm{sep}$ cut only 
(as opposed to $\Delta y$ and/or $y^*$ cuts). Thus we require that the Higgs
boson be produced between the jets, and that the minimum rapidity distance
between the Higgs boson and the two jets be larger than a minimal value,
$y_\mathrm{sep}$.

\begin{figure}[tb]
  \centering
  \epsfig{width=.8\textwidth,file=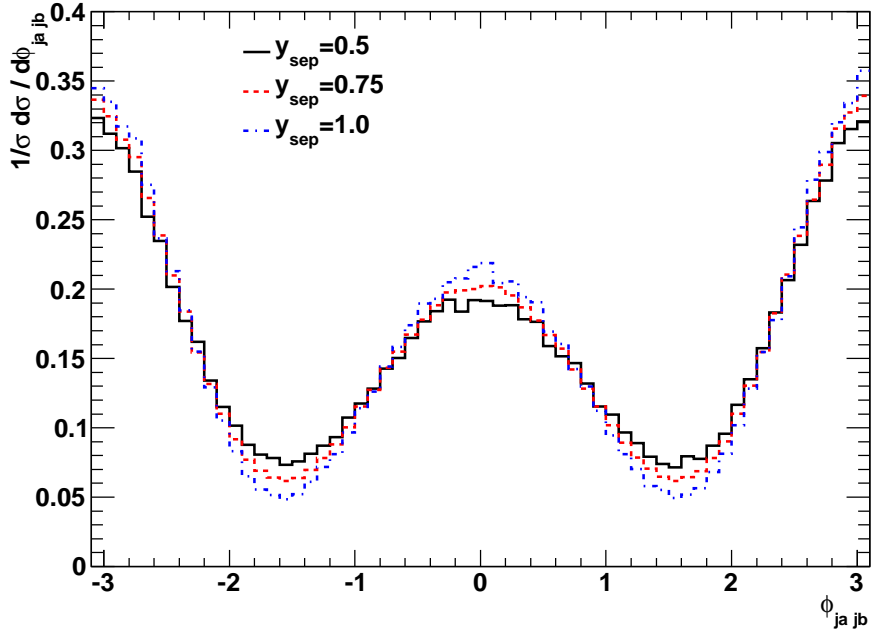}
  \vspace{-3mm}
  \caption{$(1/\sigma\ d\sigma/d\phi_{jj})$ for $hjj$-production at
    tree-level with the cuts of Eq.~\eqref{eq:multijetcuts}
    and $y_\mathrm{sep}=0.5,0.75,1.0$.}
  \label{fig:dsjjdphi2}
\end{figure}

\begin{figure}[tb]
  \centering
  \hspace{-17mm}\epsfig{width=.8\textwidth,file=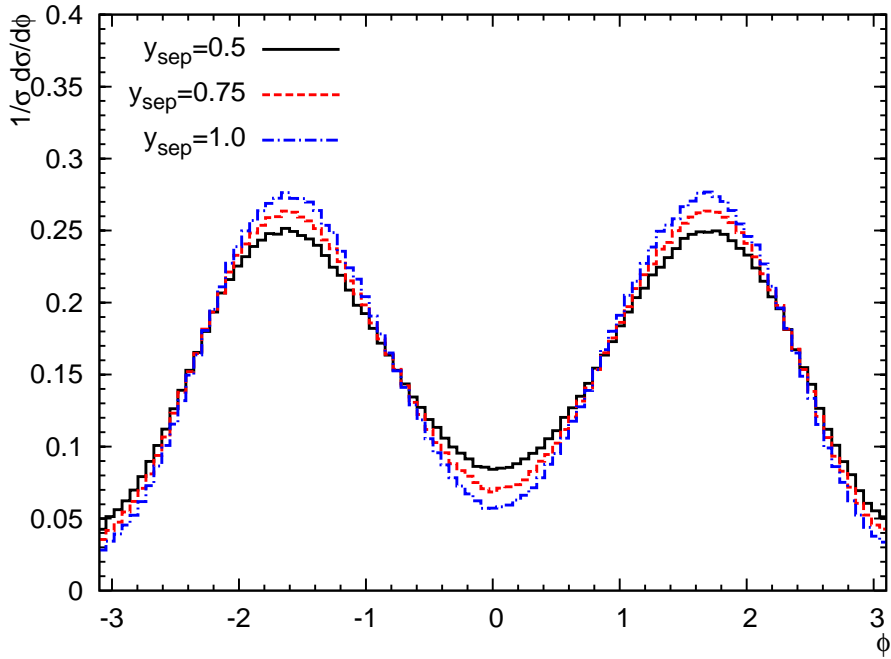}
  \vspace{-3mm}
  \caption{$(1/\sigma\ d\sigma/d\phi_{jj})$ for production of the $CP$-odd higgs boson $A$ plus two jets at
    tree-level with the cuts of Eq.~\eqref{eq:multijetcuts}
    and $y_\mathrm{sep}=0.5,0.75,1.0$. The scales were chosen as described in Section \ref{sec:parton-shower}.}
  \label{fig:dsjjdphi2CPodd}
\end{figure}

Based on the findings presented in Fig.~\ref{fig:AphivsDyys} and the
structure of the $n$-jet scattering amplitude in the MRK limit given by
Eq.~\eqref{eq:ngluonplush}, we suggest applying the following cut in
Higgs boson+multi-jet samples, when the emphasis is on extracting the
$CP$-structure of the Higgs boson coupling.
\begin{align}
  \label{eq:multijetcuts}
  p_{j\perp}>40\mbox{GeV},\ \exists j_a,j_b\in\{\mbox{jets}\}:
  y_{ja}<y_h<y_{jb},\ \forall j\in\{\mbox{jets}\}: |y_h-y_j|>y_\mathrm{sep},\ 
  |y_{j}|<4.5.
\end{align}
In Fig.~\ref{fig:dsjjdphi2} we plot $(1/\sigma d\sigma/d\phi_{jj})$, where
$\phi_{jj}$ is the azimuthal angle between the two jets (defined with the
sign according to Ref.~\cite{Hankele:2006ma}), with the cuts of
Eq.~\eqref{eq:multijetcuts} and for $y_\mathrm{sep}=0.5,0.75,1.0$. We see
that as $y_\mathrm{sep}$ is increased, the increase in $A_\phi$ noted in
Fig.~\ref{fig:AphivsDyys} is reflected in the angular distribution
approaching that of $\cos^2(\phi)$, as expected from
Eq.~\eqref{eq:ngluonplush}\footnote{Obviously, the $\cos^2(\phi)$-behavior
  of the cross section is reached only once not just the numerator is well
  approximated by that of Eq.~\eqref{eq:ngluonplush}, but also the
  $t$-channel propagator momenta have lost dependence on all but the
  transverse degrees of freedom. This is true in the MRK limit and assumed in
  Eq.~\eqref{eq:ngluonplush}, but the sub-asymptotic effects are incorporated
  in the formalism developed in
  Ref.\cite{Andersen:2008ue,Andersen:2008gc,Andersen:2009nu,Andersen:2009he}}. The cross
section and values of $A_\phi$ for the series of cuts are summarized in
For comparison we mention that with the same scale choices, but using the
standard weak boson fusion (WBF) cuts, the LO contribution to $hjj$ through
gluon fusion is 230~fb with $A_\phi=0.46$ (see
e.g.~Ref.\cite{Andersen:2008gc} for further details).
Table~\ref{tab:aphixsec}.
\begin{table}[htbp]
  \centering
  \begin{tabular}{|l|c|c|}\hline
$y_\mathrm{sep}$&$\sigma_{hjj}$&$A_\phi$\\\hline
0.5 & 743~fb & 0.36 \\\hline
0.75 & 553~fb & 0.41 \\\hline
1.0 & 403~fb & 0.45 \\\hline  
\end{tabular}
\caption{The value for the tree-level $hjj$-cross section with the cuts of
  Eq.~\eqref{eq:multijetcuts} and three values of $y_\mathrm{sep}$.}
\label{tab:aphixsec}
\end{table}
So the cuts in
Eq.~\eqref{eq:multijetcuts} with $y_\mathrm{sep}=1$ display the same strong
azimuthal correlation while almost doubling the cross section.

In Table~\ref{tab:aphixsecCPodd} we list the cross sections and values of
$A_\phi$ found for the case of a $CP$-odd coupling of SM-strength 
to a pseudo-scalar
through the effective vertex of Eq.~\eqref{eq:CAcosthetaab}. We note that the
magnitude of $A_\phi$ is almost unchanged compared to the $CP$-even case,
just the sign is reversed.
\begin{table}[htbp]
  \centering
  \begin{tabular}{|l|c|c|}\hline
$y_\mathrm{sep}$&$\sigma_{Ajj}$&$A_\phi$\\\hline
0.5 & 1730~fb & -0.36 \\\hline
0.75 & 1271~fb & -0.41 \\\hline
1.0 & 908~fb & -0.46 \\\hline  
\end{tabular}
\caption{The value for the tree-level $Ajj$-cross section with the cuts of
  Eq.~\eqref{eq:multijetcuts} and three values of $y_\mathrm{sep}$. The scales were chosen as in Section \ref{sec:parton-shower}.}
\label{tab:aphixsecCPodd}
\end{table}
In the remainder of this study we will focus on the case of the $CP$-even
Standard Model Higgs boson. However, it is clear that the cuts and analysis
developed in the following will be well suited for the extraction of the
$CP$-parameters, even in the case of a mixture of $CP$-even and $CP$-odd
couplings.

\boldmath
\subsection{Extracting the $CP$-Structure with a Fit}
\unboldmath
\label{sec:extracting-cp-struct}
In experimental data, there will of course be a non-negligible background 
to the gluon fusion $hjj$-signal. While the background is predominantly 
flat in $\phi$~\cite{Klamke:2007cu,klamke08,klamke09}, a flat pedestal 
in the distribution $(d\sigma/d\phi)$ does not
leave $A_\phi$ invariant. We will therefore introduce a different measure of
the azimuthal correlation by fitting the distribution to the following form
\begin{align}
  \label{eq:fitform}
  f(\phi)\ =\ A\ +\ \frac{\pi}{2}\, B\ \cos(2\phi+D)\ +\ C \phi^4,
\end{align}
where the parameters $A,B,C,D$ are fitted. The $\phi^4$-term is found to
describe well the impact of the sub-asymptotic angular dependence. In the
true asymptotic form of the scattering with a $CP$-even coupling and no background contribution,
$A=\frac \pi 2 B$, $D=C=0$. For $C=D=0$ one has $A_\phi = B/A$. 
A $CP$-odd coupling would set $D=\pi/2$. The size of $B$ (measured
in fb/rad) is a direct measure of the ability of the cuts and definition of
angle in resolving the azimuthal correlation induced by the $CP$-nature of
the Higgs boson coupling. Since in the current study we are interested in the
SM Higgs couplings only, all the fits return a value for $D$ consistent with
0. The quantity $B/A$ measures the quality of the given sample for extracting
the $CP$-properties. The values for $B/A$ obtained in the fits to the
$hjj$ tree-level distributions for $y_\mathrm{sep}=0.5,0.75,1$ are
$0.32,0.38,0.44$, respectively. We notice that as expected, the agreement between $A_\phi$
of Table~\ref{tab:aphixsec} and $B/A$ improves for increasing
$y_\mathrm{sep}$.

\subsection{Higgs Boson Production in Association with Three Jets}
\label{sec:higgs-boson-prod}
We will now discuss the stability of the shape of $(1/\sigma\ d\sigma/d\phi)$
found for $hjj$ at tree level against real emission higher order corrections
within the cuts of Eq.~\eqref{eq:multijetcuts}, when $\phi$ is defined
according to Eq.~\eqref{eq:qaqb}. The results of the MRK limit indicate that
$(1/\sigma\ d\sigma/d\phi)$ should be stable, when $\phi$ is defined
according to Eq.~\eqref{eq:qaqb}, rather than just the azimuthal angle
between the two hardest jets.
\begin{table}[tbp]
  \centering
  \begin{tabular}{|l|c|c|c|}\hline
$y_\mathrm{sep}$&$\sigma_{hjjj}$&$A_\phi$&$B/A$\\\hline
0.5 & 365~fb & 0.34 & 0.30 \\\hline
0.75 & 232~fb & 0.39 & 0.34 \\\hline
1.0 & 150~fb & 0.42 & 0.41\\\hline  
\end{tabular}
\caption{Values for the tree-level $hjjj$-cross section, $A_\phi$ and the 
  fitted $B/A$ ratios with the cuts of
  Eq.~\eqref{eq:multijetcuts} and three values of $y_\mathrm{sep}$ ($\phi$
  defined according to Eq.~\eqref{eq:qaqb}). }
\label{tab:aphixsec3j}
\end{table}
The results for the tree-level $hjjj$-cross section and the extracted $A_\phi$
are listed in Table~\ref{tab:aphixsec3j}. We note that with the definition of
the azimuthal angle of the multi-jet events according to Eq.~\eqref{eq:qaqb},
for each value of $y_\mathrm{sep}$, the extracted value for $A_\phi$ changes
only very little compared to the values extracted from the two-jet case. For
comparison, we note that if the WBF cuts are applied to the two hardest jet
in the three-jet sample, the tree-level cross section for $hjjj$ is found to
be just 76~fb, with $A_\phi=0.37$ (a change of 0.09 units), when $\phi$ is the
azimuthal angle between the two hardest jets.

The differences in the distribution $(1/\sigma\ d\sigma/d\phi)$ (with $\phi$
defined according to Eq.~\eqref{eq:qaqb}) in going from the two to three-jet
production is less than the differences between each neighboring curve in
Fig.~\ref{fig:dsjjdphi2}, corresponding to varying $y_\mathrm{sep}$. The
stability is a direct result of the definition of the angle according to
Eq.~\eqref{eq:qaqb}, designed from the insight obtained by analyzing the MRK
limit of the amplitudes, and optimized to reflect directly the $CP$-structure
of the Higgs boson coupling. This stability in the reconstruction and
extraction of the azimuthal angular dependence is encouraging for the
possibility of determining the $CP$-properties of the Higgs boson coupling
from LHC data. 

In the Section~\ref{sec:corr-mutli-jet} we will inspect the stability of this
azimuthal angular dependence in two models for all-order perturbative
corrections: 1) the Parton Shower model as implemented in {\sc Herwig++},
supplemented with the tree-level $hjj$-amplitudes as described in
Ref.\cite{DelDuca:2006hk}, 2) a model for both the real and virtual
corrections, based on the MRK limit of amplitudes, as discussed in
Refs.\cite{Andersen:2008ue,Andersen:2008gc}. These studies will probe the
sensitivity to the azimuthal correlation against soft radiation, which is not
picked up in the visible jets, and (especially in the second case) further
hard radiation. We will choose $y_\mathrm{sep}=0.75$ - the specific value of
$y_\mathrm{sep}$ does not affect the conclusions beyond the variation
indicated in Tables~\ref{tab:aphixsec}-\ref{tab:aphixsec3j}. Larger values
of the accepted cross section could be obtained by choosing a smaller value
for $y_\mathrm{sep}$.

In the remainder of this section we will test two further predictions on the
azimuthal distribution obtained from limiting behavior in
Eq.~\eqref{eq:ngluonplush} of the full $hjjj$-amplitude. According to the
cuts of Eq.~\eqref{eq:cuts}, the three-jet events will have two jets on one
side (in rapidity) of the Higgs boson, and one jet on the other. We will call
the two jets on the same side $j_a,j_b$, ordered according to hardness. The
single jet on the other side of the Higgs boson is called $j_c$. First (in
Section~\ref{sec:depend-azim-orient}), we see that the distribution in the
azimuthal angle defined according to Eq.~\eqref{eq:qaqb} for three-jet events
should be independent of e.g.~the angle between $j_a,j_b$, as long as the
azimuthal angle of the sum of these two jet vectors is fixed. Secondly,
(in Section~\ref{sec:depend-hardn-three}), we investigate the
prediction that the strong azimuthal correlation is displayed also in events
where the jets are ordered in transverse momentum as follows:
$p_{\perp,j_a}>p_{\perp,j_b}>p_{\perp,j_c}>40$~GeV, i.e.~the two hardest jets
are on the same side in rapidity of the Higgs boson, and the softest jet is
on the other side.

\subsubsection{Dependence on the Azimuthal Orientation of Two Jets}
\label{sec:depend-azim-orient}
\begin{figure}[tb]
  \centering
  \epsfig{width=.8\textwidth,file=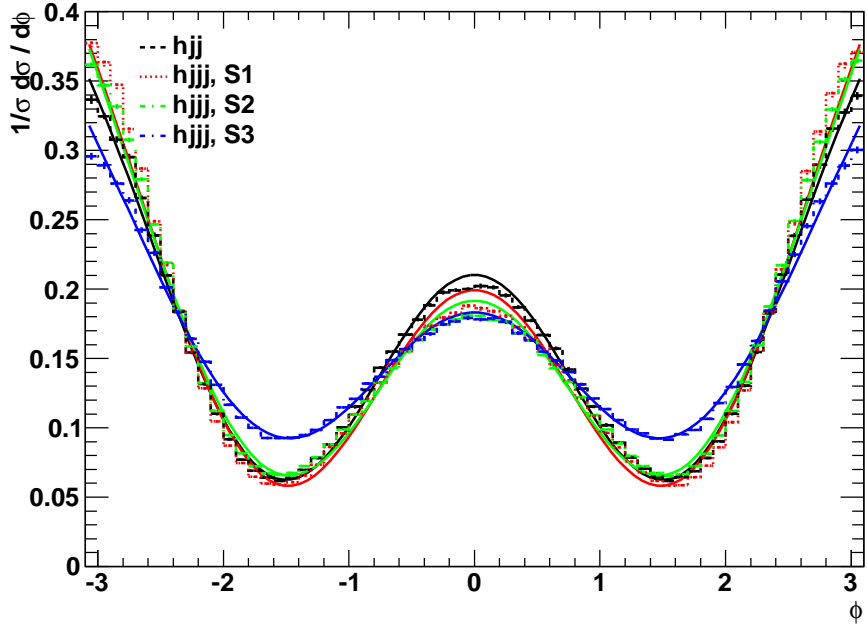}
  \vspace{-3mm}
  \caption{The angular distribution (according to Eq.~\eqref{eq:qaqb}) for
    $y_\mathrm{sep}=0.75$ for both the two-jet sample and three different
    sub-samples of the three-jet sample (all jets above 40~GeV in transverse
    momentum --- see text for further details). The
    cuts and the definition of azimuthal angle in samples of more than two
    jets ensure that the distribution is very stable against
    perturbative corrections. The lines are fits on the form of
    Eq.~\eqref{eq:fitform} to the histograms.}
  \label{fig:oneosdsdphihjjvshjjj}
\end{figure}

In Fig.~\ref{fig:oneosdsdphihjjvshjjj} we have plotted the tree-level
predictions for $(1/\sigma d\sigma/d\phi)$ for a sample of Higgs boson plus
three-jet events with-in the cuts of Eq.~\eqref{eq:multijetcuts}
($y_\mathrm{sep}=.75$), for three bins of the azimuthal angle between the two
jets on one side of the Higgs boson:
\begin{align}
  \begin{split}
    \label{eq:phiarray}
S1\ : &\   \phi_{j_a j_b}  < \pi/4\\
S2\ : &\  \pi/4 < \phi_{j_a j_b}  < 3 \pi/4\\
S3\ : &\  3 \pi/4 < \phi_{j_a j_b}  < \pi.
  \end{split}
\end{align}
For reference, we have also plotted the distribution in the azimuthal angle
in the two-jet sample. It is clear that the azimuthal dependence with the
definition of Eq.~\eqref{eq:qaqb} for the three-jet samples is extremely
stable when going to the three-jet samples $S1$ and $S2$. Only for the sample
$S3$ is there a slight change - this is because this sample is forced to have
the two jets on one side of the Higgs boson back-to-back. Since the cross
section is dominated by configurations where the jet transverse momentum is
close to the lower cut-off, the sum of the transverse momenta of the two jets
will typically be close to 0, and the square of the momentum of the
$t$-channel propagator is not dominated by its transverse momentum. In other
words, the last requirement of the MRK limit in Eq.~\eqref{eq:MRKlimit} is
violated in this sample, introducing a large dependence of the full
$t$-channel propagator on the azimuthal orientation, so the azimuthal
dependence is not described by the numerator alone. This is one reason why
the formalism of
Ref.\cite{Andersen:2008ue,Andersen:2009he,Andersen:2008gc,Andersen:2009nu}
(which implements the full momentum-dependence on the $t$-channel
propagators) offers an improvement on the BFKL
formalism\cite{Fadin:1975cb,Kuraev:1977fs,Balitsky:1978ic}. Fig.~\ref{fig:oneosdsdphihjjvshjjj}
also shows the fit according to Eq.~\eqref{eq:fitform} to each of the four
distributions. It is clear that the functional form in Eq.~\eqref{eq:fitform}
describes the distributions very well. The value for $B/A$ obtained in the
fit are $0.39,0.35,0.24$ for the samples $S1,S2,S3$, respectively, compared
with $0.34$ for the full three-jet sample with the same $y_\mathrm{sep}=0.75$.

\subsubsection{Dependence on the Hardness of the Three Jets}
\label{sec:depend-hardn-three}
Let us now turn to the case with two harder jets on one side in rapidity of
the Higgs boson, and a softer on the other side, all above 40~GeV in
transverse momentum. The reason for studying these event configurations is to
determine (using the full tree-level hard scattering matrix element) whether
in this skewed setup the scattering displays the angular correlation expected
from the analysis in Section~\ref{sec:lessons-from-he}. Specifically, we
choose the following cuts
\begin{align}
  \begin{split}
    \label{eq:multijetcutssoft}
    &p_{\perp,j_a}>p_{\perp,j_b}>p_{\perp,j_c}>40\, \mathrm{GeV},\\
    &y_{j_a,j_b}<y_h<y_{j_c}\ \mathrm{or}\  y_{j_a,j_b}>y_h>y_{j_c}\\
    &|y_h-y_j|>y_\mathrm{sep},\ |y_{j}|<4.5.
  \end{split}
\end{align}
With the standard scale-choices, the cross section passing these cuts (with
$y_\mathrm{sep}=0.75$) is 82~fb, i.e.~about $1/3$ of the three-jet rate
without the additional requirement of a special configuration of the hard
jets. Note that this is exactly the fraction expected from the MRK analysis
of Sec.~\ref{sec:lessons-from-he}: There is no preference for any of the
three jets in the event to be the softest, so in $1/3$ of the events in the
inclusive 3-jet sample it is the jet which is separated in rapidity from the
two others by the Higgs boson.

\begin{figure}[tb]
  \centering
  \epsfig{width=.8\textwidth,file=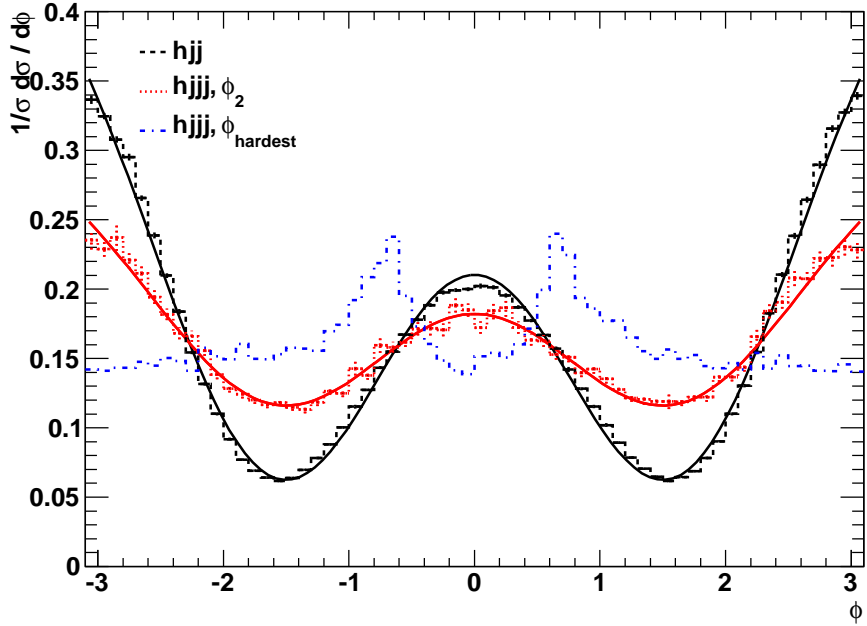}
  \vspace{-3mm}
  \caption{The angular distribution (according to Eq.~\eqref{eq:qaqb}) for
    $y_\mathrm{sep}=0.75$ for both the two-jet sample and 
    a three-jet (all above
    40~GeV in transverse momentum) sample of the two hardest jets on one side
    of the Higgs and a softer on the other (see text for further details). The
    characteristic azimuthal dependence is not reflected in the azimuthal
    angle between the two hardest jets. This dependence, induced by the
    $CP$-properties of the Higgs boson couplings, is displayed however, when
    the sum of jet vectors according to Eq.~\eqref{eq:qaqb} is
    considered. The lines are fits on the form of
    Eq.~\eqref{eq:fitform} to the histograms.
}
  \label{fig:oneosdsdphihjjvshjjj_clus}
\end{figure}

In Fig.~\ref{fig:oneosdsdphihjjvshjjj_clus} we have plotted $(1/\sigma\
d\sigma/d\phi)$ for $hjj$, and two different azimuthal distributions for the
sample of three-jet events with these selection cuts. It is immediately clear
that 1) the two hardest jets do not display even a hint of the azimuthal
correlation of the two-jet sample (the only slight enhancement is when the
two jets are pointing in the same azimuthal direction, with the collinear
divergence regularized by the jet measure ($k_t-\mathrm{jet}, R=0.6$), 2) the
vector sum of Eq.~\eqref{eq:qaqb} still displays the azimuthal correlation
induced by the Higgs boson vertex, with a value of $B/A$ from the fit to
Eq.~\eqref{eq:fitform} of $0.46$.
Therefore, if one is applying the observable of Eq.~\eqref{eq:qaqb} in the
analysis, then the events with the two hardest jets on one side of the Higgs
boson can be included and contribute to the extraction of the Higgs boson
properties. This idea is backed by the insight coming from the high energy
factorization of the hard scattering matrix element.

\section{Correlations in a Multi-Jet Environment}
\label{sec:corr-mutli-jet}

\subsection{Parton Shower}
\label{sec:parton-shower}

In order to obtain results which include higher order emissions, we used 
the parton level Monte Carlo {\sc VBFNLO} \cite{Arnold:2008rz} to 
calculate gluon fusion processes with $hjj$ signature, generating Les Houches 
Event Files, which were then used as an input for the event generator 
{\sc Herwig++} \cite{Bahr:2008pv}. The renormalization scale was set 
to $\as^4= \as(p_{ja\perp}) \as(p_{jb\perp}) \as^2(m_H)$ and the 
factorization scale was chosen as $\mu_{f}=\sqrt{p_{ja\perp}p_{jb\perp}}$. 
The computations were done using CTEQ6L1 parton distribution 
functions \cite{Pumplin:2002vw}, which is different from the choice of 
Section~\ref{sec:fixed-order-studies}. However, this has only minor 
effects and leaves the main conclusions of our findings untouched.

\begin{figure}
  \centering
  \hspace{-17mm}\epsfig{width=.8\textwidth,file=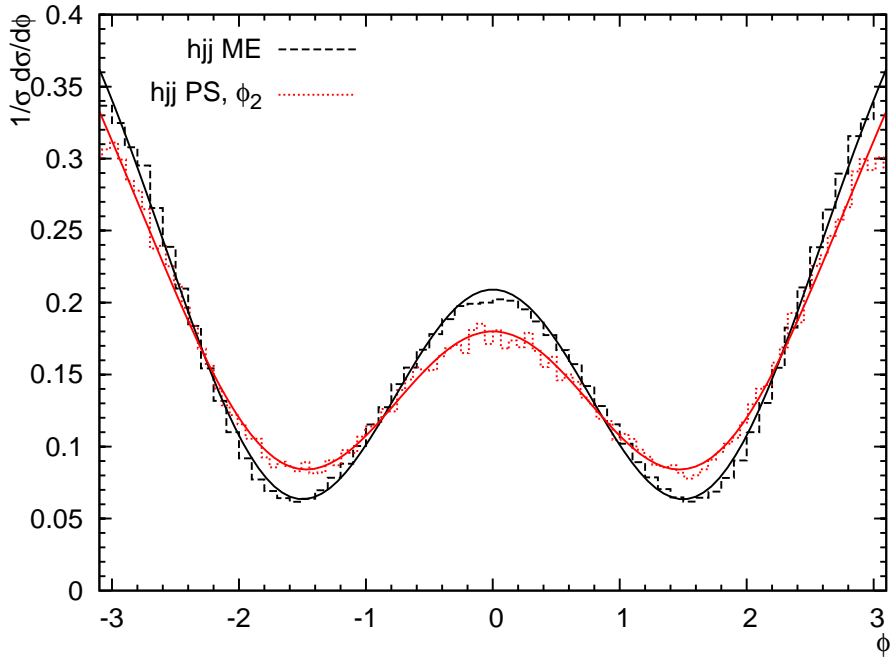}
  \vspace{-3mm}
  \caption{$(1/\sigma\ d\sigma/d\phi)$ for the $hjj$ matrix element
    calculation and the results after parton shower, both with the cuts
    of Eq.~(\ref{eq:multijetcuts}) and $y_{sep}=0.75$. $\phi$ is defined
    according to Eq.~(\ref{eq:qaqb}). The numbers for the azimuthal angle 
    asymmetry measure were extracted from the fitted function.}
  \label{fig:PShjj}
\end{figure}

We generated unweighted events using the following cuts at the parton level,
\begin{align}
  \label{eq:MEcalccuts}
  p_{j\perp}>30\,\text{GeV}, && |y_{j}|<7, && |R_{jj}|>0.3,
\end{align}
and then applied the cuts of Eq.~(\ref{eq:multijetcuts}) on the recombined jets
(which were formed using the $k_t$ jet algorithm with a cone parameter of
$R=0.6$ as in the previous section) after the shower, hadronization and
decays. An underlying event was not simulated in our studies. The choice of
rather weak cuts at the matrix element level leaves the parton shower 
evolution a great latitude, especially concerning effects altering 
rapidities of the jets.

The {\sc Herwig++} parton shower has a built in veto to forbid too hard
shower emissions. The default value for the veto scale is the highest
transverse momentum of the outgoing partons in each event at the matrix 
element level. This had to be lowered to the smallest transverse momentum 
as, otherwise, the shower produced too many hard emissions which overcame 
the characteristics of the underlying process.

\subsubsection{Higgs Boson Production in Association with Two Jets}

Fig.~\ref{fig:PShjj} shows the results after the parton shower simulation for
$y_\mathrm{sep}=0.75$. As expected, additional emissions created by the
shower lead to a decreasing amplitude of the sinusoidal modulation, from $B/A=0.38$ at $hjj$ matrix element level to $B/A=0.26$. However,
the characteristics of the azimuthal angle distribution are not significantly 
affected by the shower. Events with exactly 2 jets of $p_T>40$~GeV 
constitute 73\% of the cross section. In this exclusive 2-jet sample, 
the shower is expected to better preserve the
directions of jets which were already present in the matrix element
calculation. Indeed, for this sub-sample, the fitted azimuthal asymmetry
decreases by a somewhat smaller amount, to $B/A = 0.30$.

Since the contribution from events with more than 2 jets is
modest, it is not surprising that the angular distribution in the 2-jet
inclusive sample closely resembles the tree level result.

\subsubsection{Higgs Boson Production in Association with Three Jets}

\begin{figure}
  \centering
  \hspace{-17mm}\epsfig{width=.8\textwidth,file=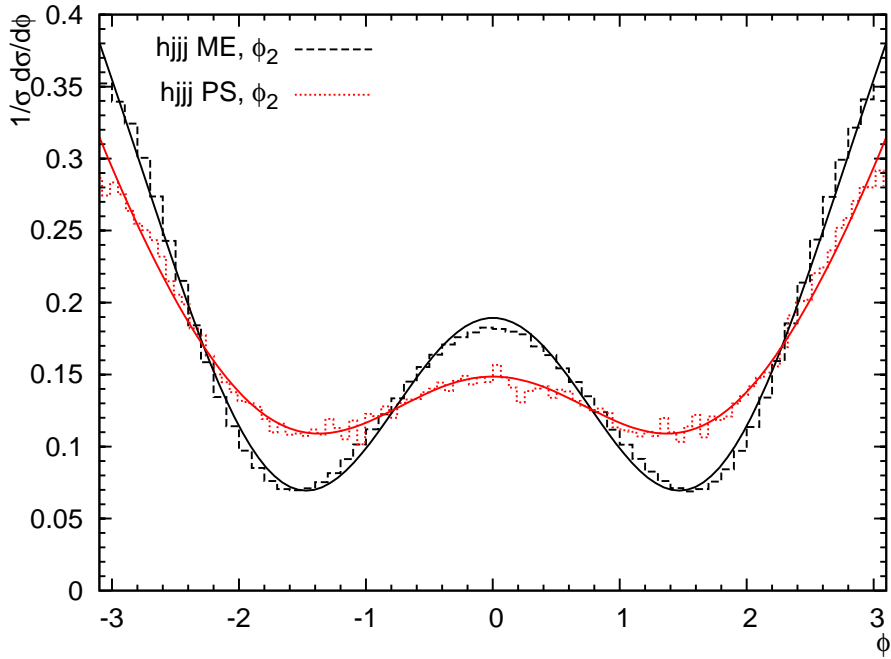}
  \vspace{-3mm}
  \caption{The azimuthal angle correlations for the $hjjj$ sample with transverse momentum of each jet above 40~GeV and $y_{sep}=0.75$. The figure shows the result of the matrix element calculation and the distribution after the parton shower simulation based on the $hjj$ matrix element calculation.}
  \label{fig:PShjjj}
\end{figure}

\begin{figure}
  \centering
  \hspace{-17mm}\epsfig{width=.8\textwidth,file=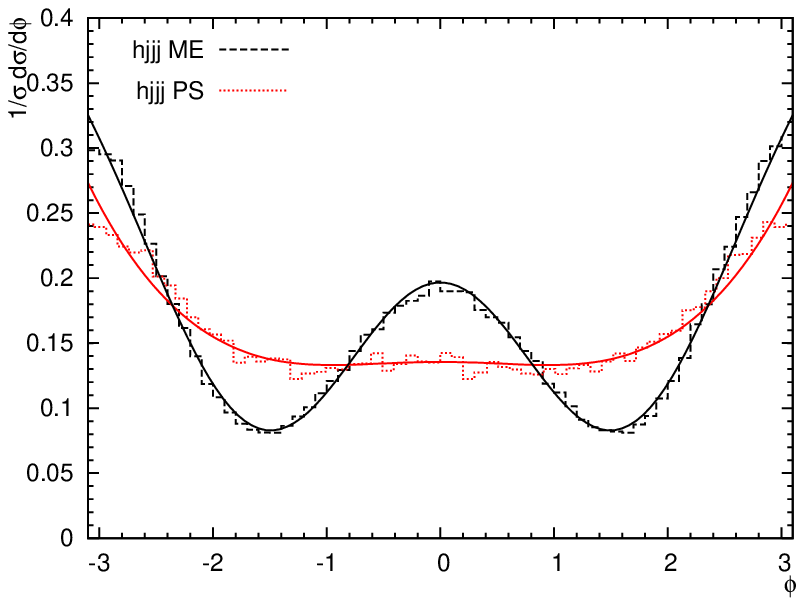}\\
  \hspace{-12.5mm}\epsfig{width=.77\textwidth,file=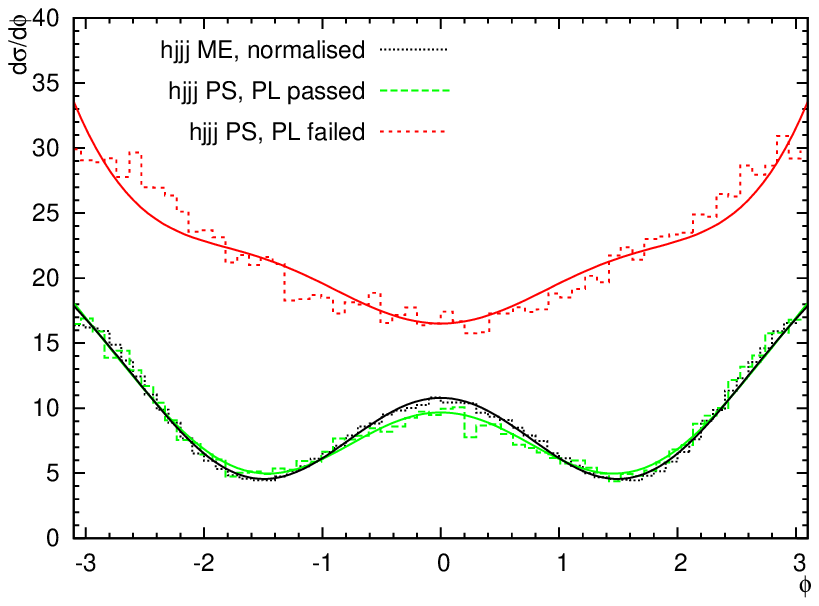}
  \vspace{-3mm}
  \caption{The angular distribution for $y_{sep}=0.75$ for the sample with
    two hard jets on one side of the Higgs Boson and a softer (still with
    $p_{\bot}>40\text{GeV}$) on the other. The two graphs in the upper plot
    depict the results of the matrix element calculation and after parton
    shower. The lower plot shows the observable for two different subsamples:
    One where the events pass the cut of Eq.~(\ref{eq:cuts}) already on
    matrix element level with just two partons (denoted with 'PL passed') and
    one where they do not ('PL failed'). The black dotted curve is the same as
    in the upper plot, but normalized to the cross section of the subsample 
    which passes the cuts at the $hjj$-parton level.}
  \label{fig:PShjjj404040}
\end{figure}

The situation changes, once the parton shower is asked exclusively for events
with three jets in the final state which, however, are still based on the 
hard $hjj$ matrix elements. In Fig.~\ref{fig:PShjjj}, the azimuthal
angle distribution for these events is plotted as the red dotted histogram
and compared to the corresponding distribution at the $hjjj$ hard matrix 
element level (black dashed histogram), which was already shown as the 
red curve in Fig.~\ref{fig:oneosdsdphihjjvshjjj_clus}. 
The difference between the results from the parton shower and that of 
the fixed order calculation is larger than in the exclusive 2-jet case. 
As the shower includes effects which go
beyond the tree-level matrix element calculation, some additional
decorrelation is expected. However, these effects of the radiation beyond
that of the tree-level are small, as shown in the previous section for
the exclusive two-jet sample. Rather, as we will show next,
a substantial part of the decorrelation is due to phase space regions 
where the parton shower completely fails to produce the azimuthal 
angle modulation inherent in the exact matrix elements. The shower
algorithm currently implemented in {\sc Herwig++} uses only azimuthally
averaged emissions, and we will in the following check the description of the
three-jet events, where at least one jet must arise from the parton shower
description.

\begin{figure}[t]
  \centering
  \hspace{-15mm}\epsfig{width=.8\textwidth,file=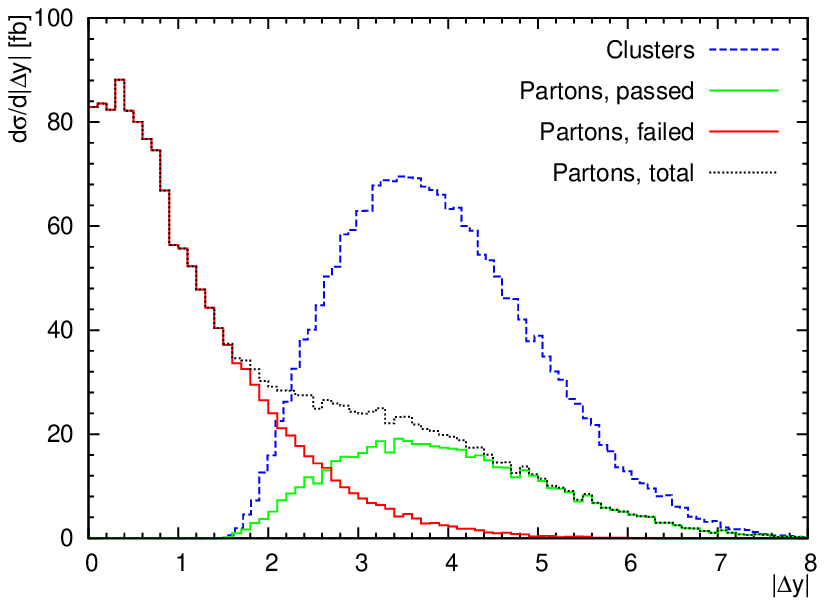}
  \vspace{-3mm}
  \caption{Absolute values of rapidity differences for $y_{sep}=0.75$ for the
    sample with two hard jets on one side of the Higgs Boson and a softer
    (with $p_{\bot}>40\text{GeV}$) on the other. The blue dashed graph shows
    the rapidity difference between the two clusters of jets on each side of
    the higgs boson, after parton shower. 
    The black dot-dashed graph shows the rapidity difference of
    the two partons at the matrix element level for the same sample of events.
    Again, this is split into two different subsamples, one where the 
    events pass the cut of Eq.~(\ref{eq:cuts}) at the parton level, 
    and one where they do not.}
  \label{fig:PShjjj404040deltay}
\end{figure}

Let us investigate in detail the kinematically restricted case from Section
\ref{sec:depend-hardn-three} with the two hardest jets $j_a$ and $j_b$ on one
side of the Higgs Boson and the third hardest jet $j_c$ on the other side. As
noted in Section~\ref{sec:depend-hardn-three}, this event configuration
corresponds to roughly a third of the full three-jet sample in the case of
the tree-level calculation. We focus on this sample, not because the parton
shower emission generates a good description of the full three-jet sample,
but because the failure in this particular sub-sample is so spectacular and
makes it easy to draw conclusions. We demand $p_{\bot j_{a,b,c}}>40$~GeV
and the rapidity cuts of
Eq.~(\ref{eq:multijetcuts}). The results are shown in
Fig.~\ref{fig:PShjjj404040}(a) as the red dash-dotted histogram, while
the black dashed histogram depicts the $hjjj$ fixed order
calculation on parton level. Unlike the cases which were examined before,
the azimuthal characteristics of the parton level appear to be 
completely lost and the fit returns a value of $B/A=0.01$.
The analysis was repeated with the transverse momentum
of the softest jet lowered: $p_{\bot j_{a,b}}>40\text{GeV}$, $p_{\bot
  j_{c}}>20\text{GeV}$, in order to investigate if this hierarchy in
transverse scales would help the description obtained within the parton
shower. However, this did not change the azimuthal distributions noticeably,
neither for the results obtained with tree-level $hjjj$ matrix elements, 
nor for the calculation with $hjj$+parton shower.

The question why the parton shower approximation fails to describe the $hjjj$
matrix element in this special setup can be answered by having a closer look
at the hard matrix element description underlying the events. One would 
expect the description from
the parton shower to agree better with the one arrived at using the full
three-jet matrix element in cases where the partonic $hjj$-configurations
already fulfil the cuts in Eq.~\eqref{eq:multijetcuts} and the two same side
jets from the shower can be produced by final state collinear splittings.
This is indeed the case, as shown by the green dashed histogram in 
Fig.~\ref{fig:PShjjj404040}(b). These events alone yield $B/A=0.23$
while their complement, where the original $hjj$ parton configurations
fail the cuts of Eq.~\eqref{eq:multijetcuts}, have lost any azimuthal 
angle correlation and produce $B/A=-0.07$.

The clearest difference between events which pass at $hjj$ parton level
and those that fail can be seen by comparing rapidity separations of 
partons and produced clusters of jets. Fig.~\ref{fig:PShjjj404040deltay} 
shows the rapidity difference 
\begin{align}
|\Delta y|=|y_{c_a} - y_{c_b}|
\end{align}
of the two jet clusters $c_a$ and $c_b$, which are present after the
recombination of hadronic final states, as the blue dashed line. For the
same sample of events, the rapidity difference of the two 
partons at matrix element level is 
plotted via the black dash-dotted histogram for all events, and this curve
is split into events wich fail the cuts at parton level (red solid histogram)
and those that pass (green dotted histogram). The sample with negligible
azimuthal angle correlation in Fig.~\ref{fig:PShjjj404040}(b)
was evolved from events which feature only a small rapidity
separation between the two outgoing partons on matrix element level and which,
therefore, have both partons on the same side of the Higgs boson. 
The parton shower does not generate the azimuthal correlation of the clusters
on opposite sides of the Higgs for such events.

As a conclusion, the question of azimuthal angle decorrelation in the setup
with two hard jets on one side of the Higgs Boson and a softer one on the 
other
is too exclusive to be handled
reliably within the parton shower approximation presently implemented
in {\sc Herwig++}: The azimuthal angle
averaged parton shower is not the right tool for the observable in this
case. It is likely, however, that the description would be improved if
spin-correlations were implemented in both the parton
shower~\cite{Knowles:1988vs,Collins:1987cp} and the interface to the matrix
element.

\subsection{Multi Parton Emissions from High Energy Factorization}
\label{sec:FKL-improved-parton-shower}

\begin{figure}[t]
  \centering
  \epsfig{width=.8\textwidth,file=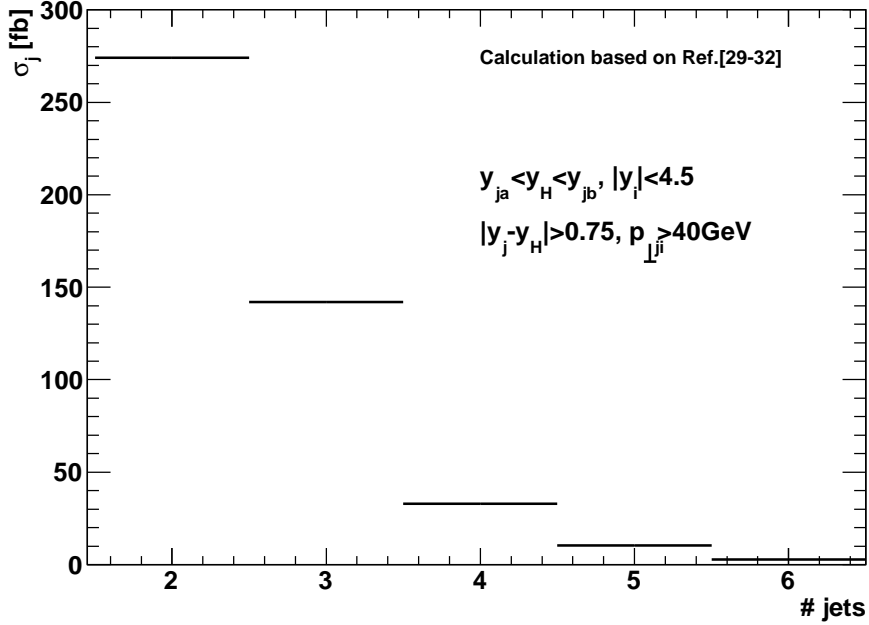}
  \vspace{-3mm}
  \caption{The exclusive jet rates within the cuts of Eq.~\eqref{eq:cuts} as
    obtained in the all-order calculation of
    Ref.\cite{Andersen:2009he,Andersen:2009nu,Andersen:2008gc,Andersen:2008ue}. All
  jets have a transverse momentum greater than 40~GeV.}
  \label{fig:resumjetrates}
\end{figure}

\begin{figure}[h!]
  \centering
  \epsfig{width=.8\textwidth,file=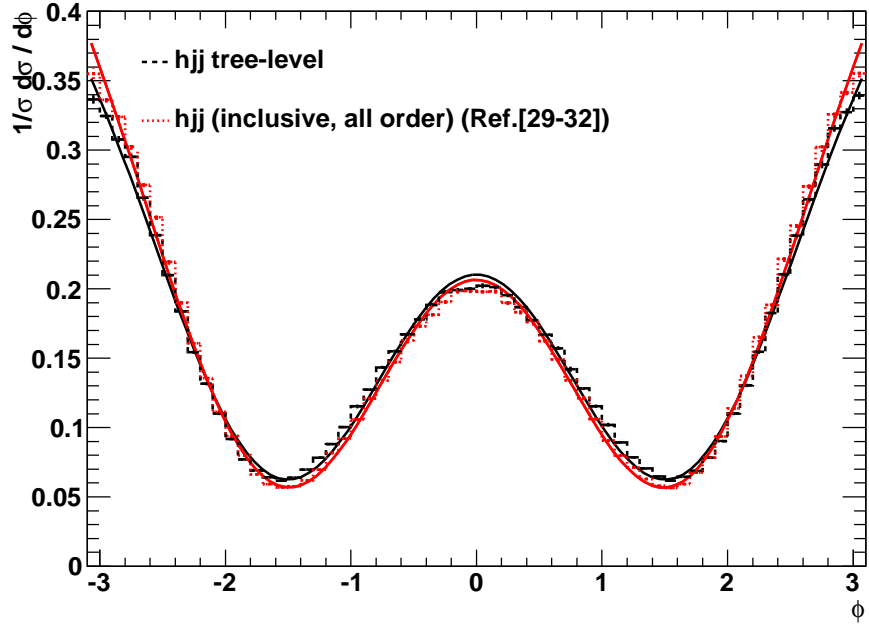}
  \vspace{-3mm}
  \caption{A comparison between the results obtained for inclusive $hjj$
    at tree-level and for the all-order approximation of
    Ref.\cite{Andersen:2009he,Andersen:2009nu,Andersen:2008gc,Andersen:2008ue}
    for the azimuthal correlation defined according to
    Eq.~\eqref{eq:qaqb}. The similarity is a testament to both the quality
    of the approximations in
    Ref.\cite{Andersen:2009he,Andersen:2009nu,Andersen:2008gc,Andersen:2008ue}
  and the stability of the observable against higher order corrections. 
    The lines are fits on the form of
    Eq.~\eqref{eq:fitform} to the histograms.
}
  \label{fig:dsdphi2_xresum}
\end{figure}

In this section we investigate the description of the production of a Higgs
boson in association with multiple jets, as described by the formalism
developed in Ref.\cite{Andersen:2008ue,Andersen:2008gc}, with the
improvements of Ref.\cite{Andersen:2009nu,Andersen:2009he} implemented. This
results in an all-order partonic estimate of the production of a Higgs boson
plus at least two jets, which is exact in the limit of large invariant mass
between all (hard) partons, i.e.~specifically in the limit where the real
radiation produces additional hard jets. The calculation is completely
exclusive in the momenta of all produced particles. We can therefore perform
a detailed investigation of e.g.~the jet activity in the events.

The all-order calculation results in a cross section within the cuts of
Eq.~\eqref{eq:multijetcuts} of 456~fb for $y_\mathrm{sep}=0.75$. This is
\emph{less} than the tree-level rate (listed in
Table~\ref{tab:aphixsec}). There is a reduction in the $hjj$ tree-level cross
section as a direct result of the veto of hard jets in the region of
0.75 units of rapidity either side of the Higgs boson. The exclusive jet
rates found within this calculation are plotted in
Fig.~\ref{fig:resumjetrates}.

The exclusive hard two-jet rate amounts to just roughly 60\% of the total
rate within these cuts, as compared to 73\% found in the study of Section~\ref{sec:parton-shower}
based on a parton shower of partonic $hjj$-states. It is therefore beneficial to ensure a solid
extraction of the $CP$-structure of the Higgs coupling also in events with
strictly more than two jets, as proposed in this study. 
In Fig.~\ref{fig:dsdphi2_xresum} we compare the differential distribution in
the azimuthal variable defined in Eq.~\eqref{eq:qaqb} for the tree-level
$hjj$-calculation, and the all-order $hjj$-calculation of
Ref.\cite{Andersen:2009he,Andersen:2009nu,Andersen:2008gc,Andersen:2008ue}. The
stability of this distribution against higher order corrections as
implemented in
Ref.\cite{Andersen:2009he,Andersen:2009nu,Andersen:2008gc,Andersen:2008ue} is
partly a result of the assumption of a dominance of the $t$-channel
pole, but this dominance is directly verified by comparing the results of the
formalism order-by-order. However, the strong correlation seen at the
tree-level (as a result of the dominance of the $t$-channel pole) could have
been spoilt in the all-order calculation by all the partons which are not
included in the momenta of the hard jets, on which the azimuthal observable
operates. However, as we see in Fig.~\ref{fig:dsdphi2_xresum}, operating only
on the momenta of the jets with a transverse momenta larger than 40~GeV still
allows for a very clear extraction of the $CP$-structure, when the observable
of Eq.~\eqref{eq:qaqb} is used.

\begin{figure}[h!]
  \centering
  \epsfig{width=.8\textwidth,file=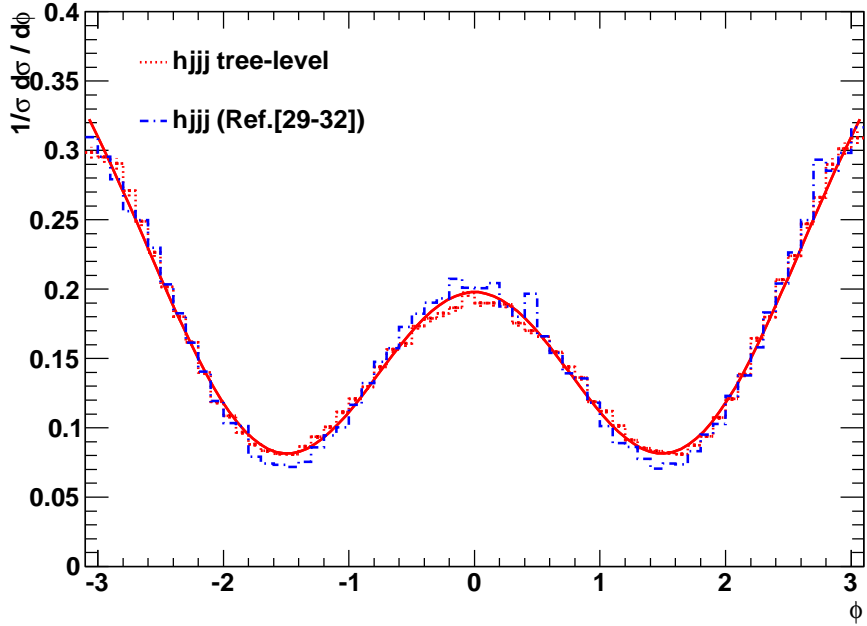}
  \vspace{-3mm}
  \caption{A comparison between the results for the azimuthal correlation
    defined according to Eq.~\eqref{eq:qaqb} obtained for the exclusive
    $hjjj$-rate with the transverse momentum of the one jet at one side of
    the Higgs boson smaller than that of either jets at the other side of the
    Higgs boson, at tree-level and for the all-order approximation of
    Ref.\cite{Andersen:2009he,Andersen:2009nu,Andersen:2008gc,Andersen:2008ue}.
    The similarity is a testament to both the quality of the approximations
    in
    Ref.\cite{Andersen:2009he,Andersen:2009nu,Andersen:2008gc,Andersen:2008ue}
    and the stability of the observable against higher order corrections. 
    The lines are fits on the form of
    Eq.~\eqref{eq:fitform} to the histograms.}
  \label{fig:dsdphi3_xresum}
\end{figure}

Finally, in Fig.~\ref{fig:dsdphi3_xresum} we compare the description of
(exclusive) three-jet events between tree-level and the resummation of
Ref.\cite{Andersen:2009he,Andersen:2009nu,Andersen:2008gc,Andersen:2008ue}. Specifically,
we compare the description within the special event sample of three hard jets
above 40~GeV transverse momentum, but with the two hardest on one side, and
the softest at the other side (in rapidity) of the Higgs boson. 

While the description of this exclusive three-jet sample differed between the
description based on the full $hjjj$ matrix element and that based on
$hjj$+parton shower, we see that the description based on
Ref.\cite{Andersen:2009he,Andersen:2009nu,Andersen:2008gc,Andersen:2008ue}
gives results which are very similar to those obtained in the calculation
based on the full three-jet tree-level matrix element. This is a testament to
both the quality of the all-order approximation applied, and the stability of
the angular observable, defined according to Eq.~\eqref{eq:qaqb}, against
higher order corrections.

\clearpage

\section{Summary and Conclusions}
\label{sec:conclusions}
Based on the analysis of the \emph{Multi-Regge-Kinematic} (MRK) limit of the
tree-level matrix elements for Higgs boson production in association with
multiple jets, we optimize the extraction of the $CP$-properties of the Higgs
boson coupling to heavy fermions in two steps. First, we have designed a set 
of cuts which enhances the sought-after behavior. Secondly, we generalize the
azimuthal angle observable used to extract the $CP$-properties of the 
couplings in Higgs boson plus dijet events to the case of three or more 
jets. This is desirable,
since roughly half of the inclusive Higgs-boson plus dijet sample will
contain more than two jets (for a 14~TeV proton-proton collider and the cuts
employed in the present analysis).

We have thoroughly tested the predictions obtained from the MRK limit against
calculations using full tree-level matrix elements, and find that the
expectations from the MRK limit are all respected by the full tree-level
calculations, and that the constructed observable leads to an extraction of
the $CP$-properties of the Higgs-boson couplings which is extremely stable
against perturbative corrections.

We have compared the predictions for both Higgs boson plus two jet and 
for three-jet
events obtained in two all-order perturbative approximations against the full
tree-level calculation. We find that the calculation based on a parton shower
evolution of the hard $hjj$-matrix element, while giving a good description 
in large areas of phase space, fails to give a satisfactory
description of the three-jet states in those phase space regions, where 
the direct correspondence of hard partons at matrix element level with 
the leading jets is lost. These phase space regions are numerically 
important. In particular, we discussed the failings
in the parton-shower description of the azimuthal correlations observed in
the full hard matrix element. Contrary to this, the all-order perturbative
calculation based on the formalism developed in
Ref.~\cite{Andersen:2008ue,Andersen:2008gc,Andersen:2009nu,Andersen:2009he}
leads to results which are in good agreement with both the exclusive two and
three-parton samples checked in this study, while giving an estimate of the
sensitivity of the observables to corrections beyond tree-level. Again, the
observable constructed in Sec.~\ref{sec:lessons-from-he} is shown to lead to
an extraction of the $CP$-properties of the Higgs boson coupling which is
extremely stable against higher order corrections.

Based on the analysis of $hjj$ and $hjjj$ at fixed order, and the
MRK-inspired all-order resummation, we conclude that, with the right choice
of observable, the analyzing power of azimuthal jet correlations in $hjj$
events for the extraction of Higgs boson $CP$-properties can be preserved
when going from a LO analysis to a more complete and sufficiently detailed
QCD description of additional parton emission. While the contribution from
hard radiation beyond tree-level is significant and can jeopardize the
extraction of the $CP$-properties, the impact of such radiation is minimized
by the use of the observable advocated in Eq.~\eqref{eq:qaqb}, which depends
on all the jet vectors of the event. In particular, also the NLO QCD
calculation of Ref.~\cite{Campbell:2006xx} should be able to provide reliable
predictions for the extraction of $CP$-properties in gluon fusion induced
$hjj$ events. Stronger decorrelation effects, which were observed in the past
when performing parton shower simulations of $hjj$
events~\cite{Odagiri:2002nd,DelDuca:2006hk} are, to a considerable extent,
due to present limitations of the parton shower programs and the use of
azimuthal observables, which are unstable against radiative corrections.

The current study also provides a clear demonstration of how the insight gained
from the structure of scattering matrix elements in the MRK limit can assist
the design of analyses at the LHC.

\subsection*{Acknowledgements}
\label{sec:acknowledgements}
DZ would like to thank the CERN TH group for financial support during the
initial stages of this study. DZ and JRA thank the organizers of the Les
Houches Workshop ``Physics at TeV Colliders'' for creating a very stimulating
environment.
This work was supported in part by the DFG
via the Graduiertenkolleg ``High Energy Physics and Particle
Astrophysics'', by the BMBF under Grant No.~05H09VKG 
(``Verbundprojekt HEP-Theorie'') 
and by the Initiative and Networking Fund of the
Helmholtz Association, contract HA-101 ("Physics at the Terascale"). 

\bibliographystyle{utphys}
\bibliography{jetpapers}

\providecommand{\href}[2]{#2}\begingroup\raggedright\begin{thebibliography}{10}

\bibitem{Rainwater:1999sd}
D.~L. Rainwater and D.~Zeppenfeld, ``{Observing $H \to W^{(*)}W^{(*)} \to e^\pm
  \mu^\mp /\!\!\!{p}_T$ in weak boson fusion with dual forward jet tagging at
  the CERN LHC},'' \href{http://dx.doi.org/10.1103/PhysRevD.60.113004}{{\em
  Phys. Rev.} {\bf D60} (1999)  113004},
\href{http://arxiv.org/abs/hep-ph/9906218}{{\tt arXiv:hep-ph/9906218}}.

\bibitem{Rainwater:1998kj}
D.~L. Rainwater, D.~Zeppenfeld, and K.~Hagiwara, ``{Searching for H
  $\rightarrow$ tau tau in weak boson fusion at the LHC},''
  \href{http://dx.doi.org/10.1103/PhysRevD.59.014037}{{\em Phys. Rev.} {\bf
  D59} (1999)  014037},
\href{http://arxiv.org/abs/hep-ph/9808468}{{\tt arXiv:hep-ph/9808468}}.

\bibitem{Plehn:1999xi}
T.~Plehn, D.~L. Rainwater, and D.~Zeppenfeld, ``{A method for identifying H
  $\rightarrow$ tau tau $\rightarrow$ e+- mu-+ missing p(T) at the CERN LHC},''
  \href{http://dx.doi.org/10.1103/PhysRevD.61.093005}{{\em Phys. Rev.} {\bf
  D61} (2000)  093005},
\href{http://arxiv.org/abs/hep-ph/9911385}{{\tt arXiv:hep-ph/9911385}}.

\bibitem{Rainwater:1997dg}
D.~L. Rainwater and D.~Zeppenfeld, ``{Searching for H $\rightarrow$ gamma gamma
  in weak boson fusion at the LHC},'' {\em JHEP} {\bf 12} (1997)  005,
\href{http://arxiv.org/abs/hep-ph/9712271}{{\tt arXiv:hep-ph/9712271}}.

\bibitem{Asai:2004ws}
S.~Asai {\em et al.}, ``{Prospects for the search for a standard model Higgs
  boson in ATLAS using vector boson fusion},''
  \href{http://dx.doi.org/10.1140/epjcd/s2003-01-010-8}{{\em Eur. Phys. J.}
  {\bf C32S2} (2004)  19--54},
\href{http://arxiv.org/abs/hep-ph/0402254}{{\tt arXiv:hep-ph/0402254}}.

\bibitem{Ball:2007zza}
{\bf CMS} Collaboration, G.~L. Bayatian {\em et al.}, ``{CMS technical design
  report, volume II: Physics performance},''
\href{http://dx.doi.org/10.1088/0954-3899/34/6/S01}{{\em J. Phys.} {\bf G34}
  (2007)  995--1579}.

\bibitem{Zeppenfeld:2000td}
D.~Zeppenfeld, R.~Kinnunen, A.~Nikitenko, and E.~Richter-Was, ``{Measuring
  Higgs boson couplings at the LHC},''
  \href{http://dx.doi.org/10.1103/PhysRevD.62.013009}{{\em Phys. Rev.} {\bf
  D62} (2000)  013009},
\href{http://arxiv.org/abs/hep-ph/0002036}{{\tt arXiv:hep-ph/0002036}}.

\bibitem{Duhrssen:2004cv}
M.~Duhrssen {\em et al.}, ``{Extracting Higgs boson couplings from LHC data},''
  \href{http://dx.doi.org/10.1103/PhysRevD.70.113009}{{\em Phys. Rev.} {\bf
  D70} (2004)  113009},
\href{http://arxiv.org/abs/hep-ph/0406323}{{\tt arXiv:hep-ph/0406323}}.

\bibitem{Lafaye:2009vr}
R.~Lafaye, T.~Plehn, M.~Rauch, D.~Zerwas, and M.~Duhrssen, ``{Measuring the
  Higgs Sector},'' \href{http://dx.doi.org/10.1088/1126-6708/2009/08/009}{{\em
  JHEP} {\bf 08} (2009)  009},
\href{http://arxiv.org/abs/0904.3866}{{\tt arXiv:0904.3866 [hep-ph]}}.

\bibitem{Kauffman:1996ix}
R.~P. Kauffman, S.~V. Desai, and D.~Risal, ``Production of a higgs boson plus
  two jets in hadronic collisions,'' {\em Phys. Rev.} {\bf D55} (1997)
  4005--4015,
\href{http://arxiv.org/abs/hep-ph/9610541}{{\tt hep-ph/9610541}}.

\bibitem{Klamke:2007cu}
G.~Klamke and D.~Zeppenfeld, ``{Higgs plus two jet production via gluon fusion
  as a signal at the CERN LHC},'' {\em JHEP} {\bf 04} (2007)  052,
\href{http://arxiv.org/abs/hep-ph/0703202}{{\tt arXiv:hep-ph/0703202}}.

\bibitem{klamke08}
G.~Kl\"amke, {\em Higgs plus 2 Jet Produktion in Gluonfusion}.
\newblock PhD thesis, University of Karlsruhe, 2008.

\bibitem{klamke09}
G.~Kl\"amke, M.~Rauch, and D.~Zeppenfeld, ``in preparation,''.

\bibitem{Kauffman:1998yg}
R.~P. Kauffman and S.~V. Desai, ``Production of a higgs pseudoscalar plus two
  jets in hadronic collisions,'' {\em Phys. Rev.} {\bf D59} (1999)  057504,
\href{http://arxiv.org/abs/hep-ph/9808286}{{\tt hep-ph/9808286}}.

\bibitem{Plehn:2001nj}
T.~Plehn, D.~L. Rainwater, and D.~Zeppenfeld, ``Determining the structure of
  {Higgs} couplings at the {LHC},'' {\em Phys. Rev. Lett.} {\bf 88} (2002)
  051801,
\href{http://arxiv.org/abs/hep-ph/0105325}{{\tt hep-ph/0105325}}.

\bibitem{Hankele:2006ma}
V.~Hankele, G.~Klamke, D.~Zeppenfeld, and T.~Figy, ``Anomalous higgs boson
  couplings in vector boson fusion at the {CERN LHC},'' {\em Phys. Rev.} {\bf
  D74} (2006)  095001,
\href{http://arxiv.org/abs/hep-ph/0609075}{{\tt hep-ph/0609075}}.

\bibitem{Odagiri:2002nd}
K.~Odagiri, ``{On azimuthal spin correlations in Higgs plus jet events at
  LHC},'' {\em JHEP} {\bf 03} (2003)  009,
\href{http://arxiv.org/abs/hep-ph/0212215}{{\tt arXiv:hep-ph/0212215}}.

\bibitem{DelDuca:2006hk}
V.~Del~Duca {\em et al.}, ``{Monte Carlo} studies of the jet activity in higgs
  + 2jet events,'' {\em JHEP} {\bf 10} (2006)  016,
\href{http://arxiv.org/abs/hep-ph/0608158}{{\tt hep-ph/0608158}}.

\bibitem{Campbell:2006xx}
J.~M. Campbell, R.~K. Ellis, and G.~Zanderighi, ``Next-to-leading order higgs +
  2 jet production via gluon fusion,'' {\em JHEP} {\bf 10} (2006)  028,
\href{http://arxiv.org/abs/hep-ph/0608194}{{\tt hep-ph/0608194}}.

\bibitem{Bahr:2008pv}
M.~Bahr {\em et al.}, ``{Herwig++ Physics and Manual},''
  \href{http://dx.doi.org/10.1140/epjc/s10052-008-0798-9}{{\em Eur. Phys. J.}
  {\bf C58} (2008)  639--707},
\href{http://arxiv.org/abs/0803.0883}{{\tt arXiv:0803.0883 [hep-ph]}}.

\bibitem{Sjostrand:2007gs}
T.~Sjostrand, S.~Mrenna, and P.~Skands, ``{A Brief Introduction to PYTHIA
  8.1},'' \href{http://dx.doi.org/10.1016/j.cpc.2008.01.036}{{\em Comput. Phys.
  Commun.} {\bf 178} (2008)  852--867},
\href{http://arxiv.org/abs/0710.3820}{{\tt arXiv:0710.3820 [hep-ph]}}.

\bibitem{Gleisberg:2008ta}
T.~Gleisberg {\em et al.}, ``{Event generation with SHERPA 1.1},''
  \href{http://dx.doi.org/10.1088/1126-6708/2009/02/007}{{\em JHEP} {\bf 02}
  (2009)  007},
\href{http://arxiv.org/abs/0811.4622}{{\tt arXiv:0811.4622 [hep-ph]}}.

\bibitem{Lonnblad:1992tz}
L.~Lonnblad, ``{ARIADNE version 4: A Program for simulation of QCD cascades
  implementing the color dipole model},''
\href{http://dx.doi.org/10.1016/0010-4655(92)90068-A}{{\em Comput. Phys.
  Commun.} {\bf 71} (1992)  15--31}.

\bibitem{Catani:2001cc}
S.~Catani, F.~Krauss, R.~Kuhn, and B.~R. Webber, ``{QCD Matrix Elements +
  Parton Showers},'' {\em JHEP} {\bf 11} (2001)  063,
\href{http://arxiv.org/abs/hep-ph/0109231}{{\tt arXiv:hep-ph/0109231}}.

\bibitem{Mangano:2006rw}
M.~L. Mangano, M.~Moretti, F.~Piccinini, and M.~Treccani, ``{Matching matrix
  elements and shower evolution for top- quark production in hadronic
  collisions},'' {\em JHEP} {\bf 01} (2007)  013,
\href{http://arxiv.org/abs/hep-ph/0611129}{{\tt arXiv:hep-ph/0611129}}.

\bibitem{Fadin:1975cb}
V.~S. Fadin, E.~A. Kuraev, and L.~N. Lipatov, ``On the {P}omeranchuk
  singularity in asymptotically free theories,''
{\em Phys. Lett.} {\bf B60} (1975)  50--52.

\bibitem{Kuraev:1976ge}
E.~A. Kuraev, L.~N. Lipatov, and V.~S. Fadin, ``Multi - {R}eggeon processes in
  the {Y}ang-{M}ills theory,''
{\em Sov. Phys. JETP} {\bf 44} (1976)  443--450.

\bibitem{Kuraev:1977fs}
E.~A. Kuraev, L.~N. Lipatov, and V.~S. Fadin, ``The {P}omeranchuk singularity
  in nonabelian gauge theories,''
{\em Sov. Phys. JETP} {\bf 45} (1977)  199--204.

\bibitem{Andersen:2008ue}
J.~R. Andersen and C.~D. White, ``{A New Framework for Multijet Predictions and
  its application to Higgs Boson production at the LHC},''
  \href{http://dx.doi.org/10.1103/PhysRevD.78.051501}{{\em Phys. Rev.} {\bf
  D78} (2008)  051501},
\href{http://arxiv.org/abs/0802.2858}{{\tt arXiv:0802.2858 [hep-ph]}}.

\bibitem{Andersen:2008gc}
J.~R. Andersen, V.~Del~Duca, and C.~D. White, ``{Higgs Boson Production in
  Association with Multiple Hard Jets},''
  \href{http://dx.doi.org/10.1088/1126-6708/2009/02/015}{{\em JHEP} {\bf 02}
  (2009)  015},
\href{http://arxiv.org/abs/0808.3696}{{\tt arXiv:0808.3696 [hep-ph]}}.

\bibitem{Andersen:2009nu}
J.~R. Andersen and J.~M. Smillie, ``{Constructing All-Order Corrections to
  Multi-Jet Rates},'' {\em JHEP} {\bf 01} (2010)  039,
\href{http://arxiv.org/abs/0908.2786}{{\tt arXiv:0908.2786 [hep-ph]}}.

\bibitem{Andersen:2009he}
J.~R. Andersen and J.~M. Smillie, ``{The Factorisation of the t-channel Pole in
  Quark-Gluon Scattering},''
\href{http://arxiv.org/abs/0910.5113}{{\tt arXiv:0910.5113 [hep-ph]}}.

\bibitem{DelDuca:2003ba}
V.~Del~Duca, W.~Kilgore, C.~Oleari, C.~R. Schmidt, and D.~Zeppenfeld,
  ``Kinematical limits on {Higgs} boson production via gluon fusion in
  association with jets,'' {\em Phys. Rev.} {\bf D67} (2003)  073003,
\href{http://arxiv.org/abs/hep-ph/0301013}{{\tt hep-ph/0301013}}.

\bibitem{DelDuca:1999ha}
V.~Del~Duca, A.~Frizzo, and F.~Maltoni, ``{Factorization of tree QCD amplitudes
  in the high-energy limit and in the collinear limit},''
  \href{http://dx.doi.org/10.1016/S0550-3213(99)00657-4}{{\em Nucl. Phys.} {\bf
  B568} (2000)  211--262},
\href{http://arxiv.org/abs/hep-ph/9909464}{{\tt arXiv:hep-ph/9909464}}.

\bibitem{Cacciari:2005hq}
M.~Cacciari and G.~P. Salam, ``Dispelling the n**3 myth for the k(t)
  jet-finder,'' {\em Phys. Lett.} {\bf B641} (2006)  57--61,
\href{http://arxiv.org/abs/hep-ph/0512210}{{\tt hep-ph/0512210}}.

\bibitem{Martin:2009iq}
A.~D. Martin, W.~J. Stirling, R.~S. Thorne, and G.~Watt, ``{Parton
  distributions for the LHC},''
\href{http://arxiv.org/abs/0901.0002}{{\tt arXiv:0901.0002 [hep-ph]}}.

\bibitem{Alwall:2007st}
J.~Alwall {\em et al.}, ``{MadGraph/MadEvent} v4: The new web generation,''
  {\em JHEP} {\bf 09} (2007)  028,
\href{http://arxiv.org/abs/arXiv:0706.2334 [hep-ph]}{{\tt arXiv:0706.2334
  [hep-ph]}}.

\bibitem{Balitsky:1978ic}
I.~I. Balitsky and L.~N. Lipatov, ``The {P}omeranchuk singularity in quantum
  chromodynamics,''
{\em Sov. J. Nucl. Phys.} {\bf 28} (1978)  822--829.

\bibitem{Arnold:2008rz}
K.~Arnold {\em et al.}, ``{VBFNLO: A parton level Monte Carlo for processes
  with electroweak bosons},''
  \href{http://dx.doi.org/10.1016/j.cpc.2009.03.006}{{\em Comput. Phys.
  Commun.} {\bf 180} (2009)  1661--1670},
\href{http://arxiv.org/abs/0811.4559}{{\tt arXiv:0811.4559 [hep-ph]}}.

\bibitem{Pumplin:2002vw}
J.~Pumplin {\em et al.}, ``{New generation of parton distributions with
  uncertainties from global QCD analysis},'' {\em JHEP} {\bf 07} (2002)  012,
\href{http://arxiv.org/abs/hep-ph/0201195}{{\tt arXiv:hep-ph/0201195}}.

\bibitem{Knowles:1988vs}
I.~G. Knowles, ``Spin correlations in parton - parton scattering,''
{\em Nucl. Phys.} {\bf B310} (1988)  571.

\bibitem{Collins:1987cp}
J.~C. Collins, ``{Spin correlations in Monte Carlo event generators},''
\href{http://dx.doi.org/10.1016/0550-3213(88)90654-2}{{\em Nucl. Phys.} {\bf
  B304} (1988)  794}.

\end{thebibliography}\endgroup

\end{document}